\documentclass[fleqn,twoside]{article}
\usepackage{espcrc2}
\usepackage{epsfig}

\usepackage{graphicx}
\usepackage[figuresright]{rotating}

\newcommand{\beq}{\begin{equation}}
\newcommand{\eeq}{\end{equation}}
\def\eqref#1{(\ref{#1})}

\catcode`\@=11 
\def\lsim{\mathrel{\mathpalette\@versim<}}
\def\gsim{\mathrel{\mathpalette\@versim>}} 
\def\@versim#1#2{\vcenter{\offinterlineskip
        \ialign{$\m@th#1\hfil##\hfil$\crcr#2\crcr\sim\crcr } }}
\catcode`\@=12 

\newcommand{\AmS}{{\protect\the\textfont2
  A\kern-.1667em\lower.5ex\hbox{M}\kern-.125emS}}

\title{Towards the resolution of the $e^+e^- \to \bar N N$ 
puzzle\thanks{Invited talk at the Int. Workshop 
``Light-cone Physics: Particles and Strings", Trento, Sept. 3-11, 2001.}
}

\author{Marek Karliner\address{School of Physics and Astronomy,\\
Raymond and Beverly Sackler Faculty of Exact Sciences\\
Tel Aviv University, Tel Aviv, Israel}%
        \thanks{e-mail: \tt marek@proton.tau.ac.il}}
       
\begin{document}

\begin{abstract}
We discuss the puzzling experimental results on baryon-antibaryon
production in $e^+e^-$ annihilation close to the threshold, in particular
the fact that $\sigma(e^+e^- \to \bar n n) \gsim \sigma(e^+e^- \to \bar p
p)$. We discuss an interpretation in terms of a two-step process, via an
intermediate coherent isovector state serving as an intermediary between
$e^+e^-$ and the baryon-antibaryon system. We provide evidence that the
isovector channel dominates both $e^+e^- \to \hbox{pions}$ and from $\bar
N N$ annihilation at rest, and show that the observed ratio of
$\sigma(e^+e^- \to \bar n n) / \sigma(e^+e^- \to \bar p p)$ can be
understood quantitatively in this picture.
\vspace{1pc}
\end{abstract}

\maketitle

Experimental data from the FENICE collaboration 
\cite{Antonelli:1998fv}
indicate that 
$\sigma(e^+ e^- \to {\bar n} n)$
is relatively large close to threshold. Their
data may be compared with earlier data on $e^+ e^- \to 
{\bar p} p$ \cite{Antonelli:1994kq,Bisello:1990rf}
 and also with data on the time-reversed reaction 
${\bar p} p \to e^+ e^-$, for which more precise data are 
available \cite{Bardin:1994am}.
\begin{figure}[htb]
\centerline{\epsfig{file=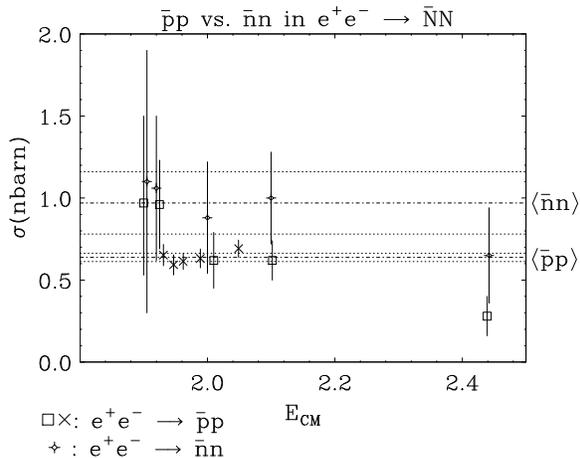,width=6.0cm,angle=90}}
\caption{\it Comparison of the cross sections for $e^+ e^- \to \bar n n$
and $\bar p p$ in the threshold region $E_{CM} \sim 2$~GeV. In the case of 
$e^+ e^- \to \bar p p$, the direct-channel data are combined with the data
for the time-reversed reaction $\bar p p \to e^+ e^-$
(marked by $\times$).
The dash-dotted and dotted lines denote the average and 1-$\sigma$ error
bars, respectively, for the $\bar p p$ and $\bar n n$ data sets.}
\label{fig:nnbar}
\end{figure}

\begin{figure}[thb] 
\centering
\centerline{\epsfig{file=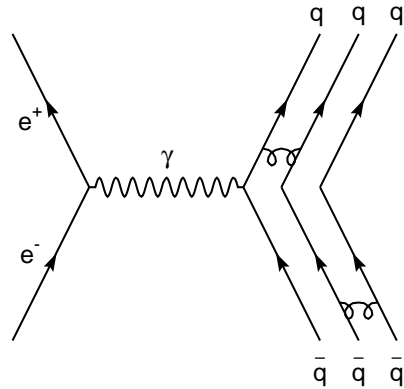,width=6.5cm}}
\caption{\it Feynman diagram corresponding to the naive perturbative
description for
$e^+e^- \,\rightarrow\, \bar N N$.}
\label{ee-fig}
\end{figure}

\begin{figure}[htb] 
\centering
\centerline{\epsfig{file=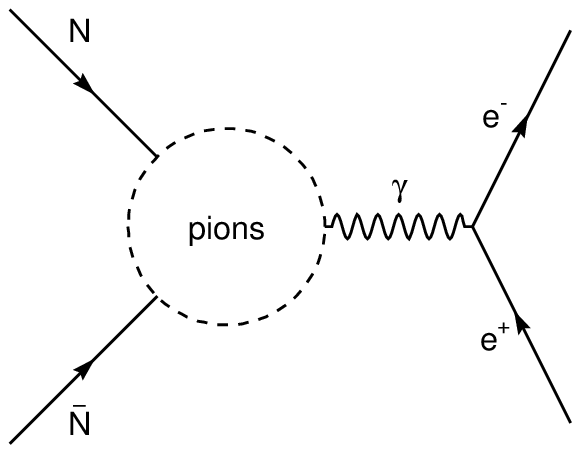,width=6.0cm}}
\caption{The time-reversed reaction, i.e.
$\bar N N  \,\rightarrow\,e^+e^-$, as
a two-stage process, where the nucleons first annihilate into
pions, which then couple to a a timelike photon to produce the
$e^+e^-$ pair.}
\label{pions1-fig}
\end{figure}

As seen
 in Fig.~\ref{fig:nnbar}, the combined data indicate 
that $\sigma(e^+ e^- \to {\bar n} n) / \sigma(e^+ e^- \to {\bar p} p) 
\gsim 1$ when $E_{CM} \sim 2$ GeV. Averaging over the available data on 
both the direct and time-reversed reactions, which are very consistent, 
and ignoring any possible variation with energy, we find:
\begin{equation}
{\sigma(e^+ e^- \to {\bar p} p) \over \sigma(e^+ e^- \to {\bar n} n)} \; = 
0.66^{+0.16}_{-0.11}
\label{exptalratio}
\end{equation}
In other words, {\em close to the $\bar N N$ threshold,
the timelike form factor of the proton is somewhat smaller
than that of the neutron!}

The fact that this ratio is less than unity requires confirmation, but 
even equal cross sections for $e^+ e^- \to {\bar p} p$ and $e^+ e^- \to 
{\bar n} n$ would be quite surprising, in view of the 
the conventional perturbative
picture of baryon-antibaryon production in
$e^+e^-$ annihilation, as shown in Fig.~\ref{ee-fig}.

We recall that the ratio of the cross sections for the corresponding
$t$-channel processes $e p (n) \to e p (n)$ should
be infinite at zero momentum transfer, where the form factor simply
measures the total proton and neutron charges, corresponding to a coherent
sum over the electromagnetic charges of their constituent quarks. It is
also believed that the ratio (\ref{exptalratio}) should be large at high
momentum transfers. In a naive perturbative description of $e^+e^-$
annihilation into baryons, the virtual time-like photon first makes a
`primary' $\bar q q$ pair, which is then dressed by two additional
quark-antiquark pairs that pop out of the vacuum. This dressing is thought
to be a perturbative QCD process at high momentum transfers, which does
not distinguish between the $u$ and $d$ quarks, since gluon couplings are
flavor-blind. Thus, in this conventional perturbative picture, the only
difference between the production rates of proton and neutron is through
the different electric charges of the primary $\bar q q$ pairs. The total
perturbative cross section is obtained by superposing the amplitudes with
different primary $\bar q q$ pairs and squaring the result:
\begin{equation}
\sigma (e^+e^- \,\rightarrow\, \bar N N)
\propto \displaystyle \left\vert \sum_{q\in N} Q_q a^N_q(s)
\right\vert^2,
\label{QED-NN}
\end{equation}
where $a^N_q(s)$ denotes the amplitude at $E^2_{CM}=s$ for making the
baryon $N$ with a given primary flavor $q$, which is determined by the
baryon wave functions.

Since the wave functions of the baryon octet have a mixed symmetry, the
amplitudes $a^N_q(s)$ tend to be highly asymmetric in specific models. For
example, in the Chernyak-Zhitnitsky proton wave
function~\cite{Chernyak:1984ej}, the $u$ quark dominates, i.e., $a^p_u =
{\cal O}(1)$, $a^p_d \ll 1$ and similarly $a^n_d = {\cal O}(1)$, $a^n_u
\ll 1$. In such a limiting case we have
\begin{equation}
{\sigma (e^+e^- \,\rightarrow\, \bar p p)
\over
\sigma (e^+e^- \,\rightarrow\, \bar n n) }
\quad \longrightarrow \quad
{Q_u^2\over Q_d^2} = 4.
\label{PT-NN}
\end{equation}
While this is an extreme case, on general grounds we expect that the $u$
contribution dominates in the proton and the $d$ in the neutron, so that
$\sigma ( e^+e^- \,\rightarrow\, \bar p p)/
 \sigma ( e^+e^- \,\rightarrow\, \bar n n) \gg 1$ at large momentum 
transfers. 

We find it puzzling that the experimental ratio (\ref{exptalratio}) is
apparently below unity when $E^2_{CM}=s \sim 4$~GeV$^2$, whereas the ratio
should be much larger than unity at both larger (timelike) and smaller
(spacelike) momentum transfers.  Clearly, the mechanism at work here is
qualitatively different from those responsible for the above intuition.

The lack of a conventional theoretical explanation is part of the
motivation for the proposed new asymmetrical $e^-e^+$ high-statistics
collider at SLAC for the regime $1.4 < \sqrt{s} < 2.5$ GeV \cite{LOI}.
This machine will yield high-precision data on baryon production in
$e^-e^+$ annihilation at threshold, providing a check on the FENICE data
and an accurate benchmark for testing possible theoretical explanations.

The first thing one must realize is that even though $q^2 \gsim 4 m_N^2
\gg \Lambda_{QCD}^2$, the process is highly nonperturbative. This is
because the `extra' kinetic energy available to the quarks is very small.

Our approach \cite{Karliner:2001ut} to this puzzle is based on thinking
about the time-reversed processes: ${\bar N} N \to e^+ e^-$. 
These may be
viewed as two-step processes, with a coherent meson state serving as an
intermediate state, as shown schematically in Fig.~\ref{pions1-fig}.

One possible motivation for this picture might be
provided by the Skyrme model~\cite{Skyrme:1961vq,Skyrme:1962vh}, according
to which baryons appear as solitons in a purely bosonic chiral Lagrangian.
This model is formally justified as a low-energy approximation to
large-$N_c$ QCD~\cite{Witten:1979kh,Adkins:1983ya}, and is known to
provide a good description of many low-energy properties of baryons:
see~\cite{Zahed:1986qz,Schechter:1999hg} for reviews.
Skyrmion-anti-Skyrmion annihilation
provides~\cite{Verbaarschot:1987rj}-\cite{amado} a fairly accurate
description of low-energy baryon-antibaryon annihilation. Just after the
Skyrmion and anti-Skyrmion touch, they `unravel' each other, and a
coherent classical pion wave emerges as a burst that takes away energy and
baryon number as quickly as causality permits. A specific parametrization
of the initial pion configuration is~\cite{Halasz:2001ye}:
\begin{equation}
F ( r, t = 0) \; = h \,{r \over r^2 + a^2} e^{-{r /a}}\ ,
\label{Amado}
\end{equation}
where $F$ is the profile of the chiral field, 
$U=\exp[\,i\,\tau\cdot\hat r\, F(r,t)\,]$,
$a$ is a range parameter, $h$ is chosen so that the total energy is
that of the $\bar N N$ pair, and the form of $F$ guarantees that the pion
configuration has zero net baryon number. This crude model has been
shown~\cite{Halasz:2001ye} to reproduce satisfactorily the inclusive
single-pion spectrum in $\bar p p$ annihilation at rest and the branching
ratios for multi-pion final states. 

The details of this specific configuration are unimportant for our
purposes: what is important is that the data are not inconsistent with
such a model. Indeed, although the Skyrme model provides some motivation
for our approach, it is not even essential for our purpose. What is
important is that {\it a single intermediate state should dominate} the
two-step ${\bar N} N \to e^+ e^-$ process. This could, for example,
equally well be a single intermediate $J^{PC} = 1^{--}$ resonant meson
state.

To be more precise, since $\bar N N$ annihilation is a strong-interaction
process, one must consider separately the $I = 1$ and $I = 0$ channels.  
Accurately stated, our key assumption is that both of these channels are
dominated by single states. These might be some excited $\rho^*$ and
$\omega^*$ mesons, for example, just as well as coherent pion
configurations (\ref{Amado}) with $I = 1$ and $I = 0$.

With this picture in mind, we write the $I = 1,0$ ${\bar N} N \to e^+ e^-$
annihilation amplitudes as $A_1, e^{i \alpha} A_0$, where the overall
phase is irrelevant, $A_1$ and $A_0$ are relatively real, and $\alpha$ is
the relative phase between the $I = 1$ and $I = 0$ amplitudes. We then
have
\begin{equation}
f \equiv {\sigma(e^+ e^- \to {\bar p} p) \over \sigma(e^+ e^- \to {\bar n} 
n)}
 = \left\vert {A_1 + e^{i \alpha} A_0 \over A_1 - e^{i \alpha} 
A_0}\right\vert^2.
\label{ratio}
\end{equation}
It is apparent from (\ref{ratio}) that $\sigma(e^+ e^- \to {\bar n} n) /
\sigma(e^+ e^- \to {\bar p} p) \sim 1$ if either $A_1 \gg A_0$ or {\it
vice versa}.

Remarkably, {\it there is evidence from both $e^+ e^-$ and $\bar N N$
annihilations that $I = I$ final states dominate by large factors}. 

\begin{figure}
\centerline{\epsfig{file=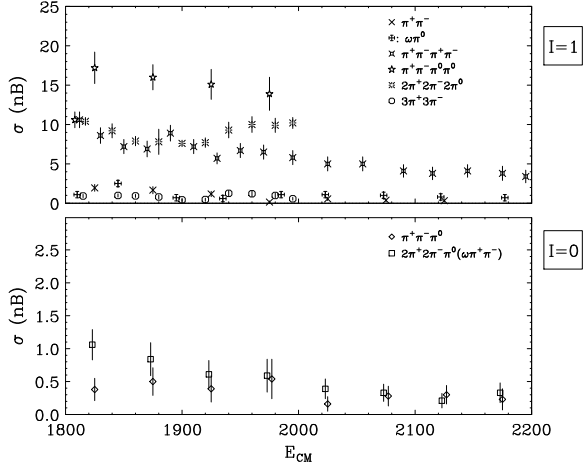,width=6.2cm,angle=90}}
\caption{\it Cross sections for $e^+ e^- \to$ multi-pion final 
states, for $E_{CM} \sim 2$~GeV \cite{eidelman}.
We note the dominance of $I = 1$ final 
states with even numbers of pions by about an order of magnitude over 
$I = 0$ states with an odd number of pions.}
\label{fig:pions}
\end{figure}
The clearest evidence comes from $e^+ e^- \to n \pi$, where it is found by
measuring final states with even and odd numbers of pions
respectively
that
\begin{equation}
{\sigma(e^+ e^- \to (2m) \pi) \over \sigma (e^+ e^- \to (2m + 1) \pi)} 
\sim 9 \; \; {\rm for} \;\; E_{CM} \sim 2 \; {\rm GeV},
\label{eeratio}
\end{equation}
as seen
in Fig.~\ref{fig:pions}.\footnote{The five-pion final state
is predominantly 
$\omega \pi\pi$. The cross section in Fig.~\ref{fig:pions}
corresponds to the final state
$\omega \pi^+ \pi^-$;
for the total $\omega \pi\pi$ one should multiply it by 1.5
\cite{eidelman}.}
At these energies, we expect most final states
created by $e^+ e^- \to \bar s s$ to contain $K \bar K$ pairs, so that
(\ref{eeratio})  corresponds to the non-$\bar s s$ initial states we
expect to dominate in $\bar N N$ annihilation. The value (\ref{eeratio})
is similar to that found at lower energies, where $\Gamma(\rho \to e^+
e^-) \sim 9 \times \Gamma(\omega \to e^+ e^-)$, in agreement with naive
quark models. The fact that the ratio $\sigma ( I = 1) / \sigma ( I = 0)$
continues to be large at higher energies is consistent with ideas of
generalized vector meson dominance.
The data from ${\bar N} N$ annihilations are less clear.  Theoretically,
there are various calculations and other suggestions that the cross
sections for ${\bar N} N$ annihilations into $I = 1$ and $I = 0$ may be
similar. Experimentally, several initial states contribute, including
$^1$S$_0$, $^3$S/D$_1$ and various P waves, but we are interested only in
the $J^{PC} = 1^{--}$ $^3$S/D$_1$ initial states. It is in principle
possible to distinguish different initial states by comparing
annihilations in gas and liquid, as has been done in the analysis of
OZI-violating final states, but we are unaware of a comparable analysis of
multi-pion final states. Because the initial state is a mixture with
different $G = \pm 1$, it is not possible to separate $I = 1$ from $I = 0$
simply by counting pions, as was the case in $e^+ e^-$ annihilation.

The most convincing experimental information known to us comes from an 
analysis of ${\bar N} N \to {\bar K} K$. By comparing the rates for $\bar 
p p \to K^+ K^-$ and $\bar p p \to K^0 {\overline K^0}$,
it has been possible to extract~\cite{Dover:1992vj}
\begin{equation}
\left\vert { A(^3S/D_1 \to {\bar K} K)_{I = 1} \over
A(^3S/D_1 \to {\bar K} K)_{I = 0}} \right\vert^2 \sim 5 \; \; {\rm to} \; 
\; 10,
\label{NNbarratio}
\end{equation}
which is comparable to the corresponding ratio (\ref{eeratio}) in $e^+
e^-$ annihilation.  The fact that $I = 1$ dominates over $I = 0$ in the
ratio (\ref{NNbarratio}) is consistent with the hypothesis that most of
the ${\bar K} K$ final states are created by $\bar u u$ and $\bar d d$
pairs in the initial ${\bar N} N$ state, with a $\bar s s$ pair popping
out of the vacuum. The ratio (\ref{NNbarratio}) would be small if primary
$\bar s s$ pairs dominated.

We now use the experimental information on the dominance by the $I = 1$
channel in both $e^+ e^-$ (\ref{eeratio}) and $^3$S/D$_1$ annihilation
(\ref{NNbarratio}) in a quantitative analysis of the $e^+ e^- \to \bar N
N$ production $f$ ratio (\ref{ratio}). Defining $\epsilon \equiv A_0 / 
A_1$, we can rewrite (\ref{ratio}) as
\begin{equation}
f \;  = \; \left\vert {1 + e^{i \alpha} \epsilon \over 1 - e^{i \alpha}
\epsilon}\right\vert^2.
\label{newratio}
\end{equation} 
It is clear that the zero-momentum-transfer limit $\sigma (e^+ e^- \to 
\bar p p) \gg (e^+ e^- \to \bar n n)$ is obtained in the limit $\epsilon 
\to 1, \alpha \to 0$, and that the high-momentum-transfer limit $\sigma 
(e^+ e^- \to \bar p p) \sim 4 \times (e^+ e^- \to \bar n n)$ is obtained 
in the limit $\epsilon \to 1/3, \alpha \to 0$. In order to estimate 
$\epsilon = A_1 / A_0$, our assumption of dominance in each isospin 
channel by a single state 
(either a coherent multi-pion state (\ref{Amado}) or a generalized vector 
meson $V^*$) tells us that
\begin{equation}
\epsilon = \sqrt {{ \sigma (e^+ e^- \to (I = 0)) \over \sigma (e^+ e^- \to 
(I = 
1))}} \times \sqrt {{ \sigma (\bar N N \to (I = 0)) \over \sigma (\bar N N 
\to (I = 1))}}.
\label{factorization}
\end{equation}
Inserting the experimental indications (\ref{eeratio},\ref{NNbarratio}) 
into (\ref{factorization}), we estimate that
\begin{equation}
{1 \over 10} \; \lsim \; \epsilon \; \sim \; {1 \over 3}.
\label{epsilon}
\end{equation}
The top end of this range seems to us quite conservative, whereas the 
lower end surely requires more justification from $\bar N N$ annihilation 
data. In the following numerical analysis, we keep $\epsilon$ general, but 
focus extra attention on the limits $\epsilon = 1/3$ and $1/10$.
 
It is apparent from (\ref{newratio}) that $f \sim 1$ is possible for 
any value of $\epsilon$, for a restricted range of the relative 
phase $\alpha \sim \pi / 2$. However, the allowed range of $\alpha$ is 
extended if $\epsilon$ is small. It is easy to see that $f$ lies in a 
narrow range $\Delta f$ around unity if $\alpha$ falls within the
following range:
\begin{equation}
\Delta \alpha \; \simeq \; {\Delta f \over 4 \epsilon}.
\end{equation}
It is apparent that $f \sim 1$ for all values of $\alpha$ if $\epsilon$ is 
small, as suggested (\ref{epsilon}) by the available data on $e^+ e^-$ and 
$\bar N N$ annihilation. 

\begin{figure}
\centerline{\epsfig{file=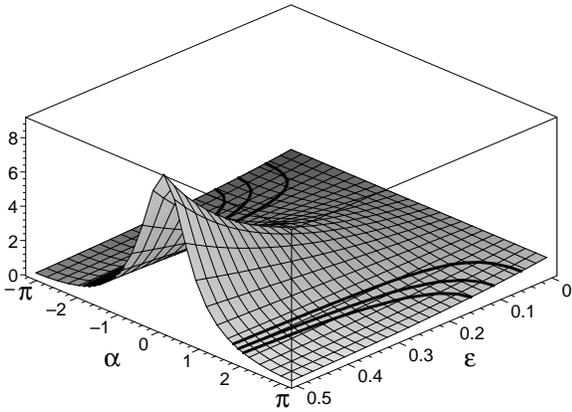,width=5.3cm,angle=270}}
\vspace{0.2in}
\caption{\it Three-dimensional plot of the cross-section ratio $f \equiv 
\sigma(e^+ e^- \to {\bar p} p) / \sigma(e^+ e^- \to {\bar n} n)$ as a 
function of $\epsilon$ and $\alpha$, indicating the region where $f$ falls 
within the range (\ref{exptalratio}).}
\label{fig:threeD}
\end{figure}

\begin{figure}[thb]
\bigskip
\centerline{\epsfig{file=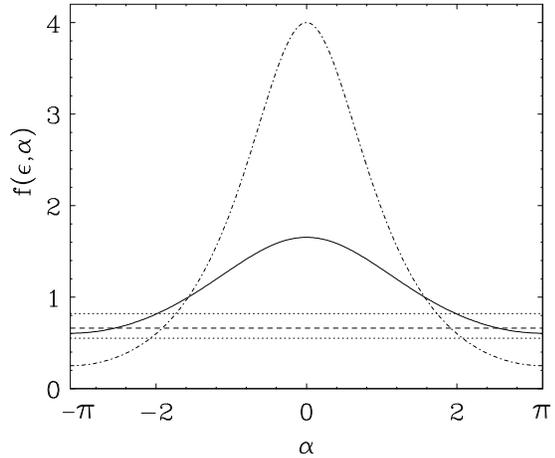,width=6.0cm,angle=90}}
\vspace{0.2in}  
\caption{\it Two-dimensional plot of the cross-section ratio $f \equiv
\sigma(e^+ e^- \to {\bar p} p) / \sigma(e^+ e^- \to {\bar n} n)$ as a   
function of $\alpha$ for $\epsilon = 1/3$ (dot-dash curve)
and $\epsilon=1/10$ (continuous curve), indicating the 
ranges where $f$ falls within (\ref{exptalratio}).}
\label{fig:twoD}
\end{figure}

The quantitative behaviour of $f$ as a function of $0 < \epsilon < 1/2$
and $- \pi < \alpha < \pi$ is shown in Fig.~\ref{fig:threeD}. Displayed
explicitly is the region of the $(\epsilon, \alpha)$ plane where $f$ falls
within the experimental range (\ref{exptalratio}). We see that this range
favours $|\alpha| > \pi / 2$, whatever the value of $\epsilon$.  
Fig.~\ref{fig:twoD} displays projections of Fig.~\ref{fig:threeD} for the
two limiting values $\epsilon = 1/3$ and $1/10$. The allowed range
(\ref{exptalratio}) of $f$ and the corresponding ranges of $\alpha$ are
also shown.

We conclude that the {\it a priori} puzzling large experimental value of
the ratio $\sigma(e^+ e^- \to {\bar n} n) / \sigma(e^+ e^- \to {\bar p}
p)$ can be understood qualitatively. This is relatively easy if the $I =
1$ amplitude dominates over the $I = 0$ amplitude, as suggested by the
available data on $e^+ e^-$ and $\bar N N$ annihilation and our assumption
of dominance by a single coherent state in each isospin channel. The
specific range (\ref{exptalratio}) can be understood quantitatively if the
$I = 1$ and $I = 0$ amplitude have a large relative phase $\alpha$. We are 
not in a position to judge the plausibility of such a large value of 
$\alpha$, from either an experimental or a theoretical point of view. It 
would be interesting to make tests of this possibility.

\begin{figure}[thb]
\centerline{\epsfig{file=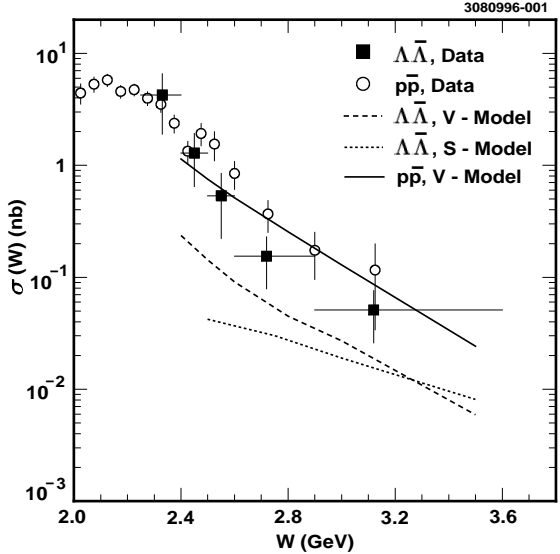,width=7.5cm}}
\caption{CLEO data
\cite{Anderson:1997ak}
for
$\sigma_{\gamma\gamma \,\rightarrow\, \Lambda\overline{\Lambda}}(W)$,
$\sigma_{\gamma\gamma \,\rightarrow\, p\overline{p}}(W)$
for $|\cos{\theta^*}| < 0.6$.
Vertical error-bars include
systematic uncertainties.
Horizontal markings indicate bin width.
S-model: scalar quark-diquark model;
V-model: vector quark-diquark model.
}
\label{cleo-gg-fig}
\end{figure}

\begin{figure}[thb] 
\centerline{\epsfig{file=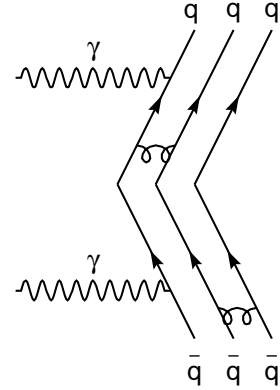,width=4cm}}
\caption{Feynman diagram corresponding to the naive perturbative
description for
$\gamma\gamma \,\rightarrow \,\bar N N$.}
\label{gg-fig}
\end{figure}
The disagreement between the naive theoretical
prediction and experiment is striking again.

We comment finally on the surprisingly large value of the ratio
$\sigma(\gamma \gamma \to \bar \Lambda \Lambda) / \sigma(\gamma \gamma \to
\bar p p)$ observed by the CLEO Collaboration \cite{Anderson:1997ak} (see
also \cite{Artuso:1994xk}-\cite{Hamasaki:1997cy} for related experimental
work). 

As shown in Fig.~\ref{cleo-gg-fig},
CLEO find that \hbox{$\sigma(\gamma\gamma \,\rightarrow\,
p\overline{p})\approx \sigma(\gamma\gamma \,\rightarrow\,
\Lambda\overline{\Lambda})$} close to threshold, which seems analogous to
FENICE result for the $\bar n n/\bar p p$ ratio. 

As illustrated by Fig.~\ref{gg-fig},
for a given quark flavor,
the perturbative amplitude for baryon-antibaryon production in the
photon-photon reaction scales like the quark charge squared, compared with
the linear dependence of the amplitudes on the quark charge in the
$e^+e^-$ case discussed earlier. 
Thus one might naively expect the ratio
$\sigma(\overline{\Lambda}\Lambda)/\sigma(\bar p p)$ to be even smaller
than the corresponding perturbative prediction for $\sigma(\bar n
n)/\sigma(\overline{p}p)$ in $e^+e^-$. 

It would be interesting to approach
this puzzle from a point of view similar to that adopted in this paper.
However, the situation in $\gamma \gamma$ collisions is more complicated,
because of the wider range of possible spin and isospin states. Also, the
information available on the isospin and spin decomposition is sparse
compared with that in $e^+e^-$ annihilation, which we used above. Data for
$\gamma\gamma \,\rightarrow\, \overline{n} n$ close to threshold might
cast light on the $\sigma(\gamma \gamma \to \bar \Lambda \Lambda) /
\sigma(\gamma \gamma \to \bar p p)$ puzzle, but are not yet available.

In Ref. \cite{Ellis:2001xc} 
we have proposed a simple model that is able to accommodate
the surprisingly large observed value of the ratio $\sigma(e^+e^- \to \bar
n n) / \sigma(e^+e^- \to \bar p p)$. Our suggestion is based on a simple
two-step approach, in which a single intermediate state with $I = 1$
dominates over $I = 0$. This dominant intermediate state could be
motivated by a Skyrmion-anti-Skyrmion picture, or could be some excited
$\rho^*$ resonance. Our model could be tested by further measurements of 
the ratios of different isospin amplitudes in $e^+ e^-$ and 
$\bar N N$ annihilation, and suggests a relatively large phase difference 
between $I = 1$ and $I = 0$ amplitudes. We look forward to more 
experimental data bearing on these issues, for example from a new 
low-energy $e^+ e^-$ collider~\cite{LOI}.

\section*{Acknowledgments}

This talk is based on Ref.~\cite{Ellis:2001xc}
and on earlier work with Stan Brodsky, summarized
in Ref.~\cite{Karliner:2001ut}.
I would like to thank the organizers of the Trento 
International Light-Cone Workshop
for their hospitality and for setting up such a well-organized meeting.
The research of one of M.K. was supported in part by a grant from the
United States-Israel Binational Science Foundation (BSF), Jerusalem,
Israel, and by the Basic Research Foundation administered by the Israel
Academy of Sciences and Humanities.

\end{document}